\PassOptionsToPackage{usenames,dvipsnames,table}{xcolor}
\documentclass[twocolumn,aps,prd,longbibliography,nofootinbib,longbibliography,notitlepage]{revtex4-1}
\usepackage{graphicx,epsfig,epstopdf}
\usepackage{amsmath,amssymb,latexsym,bm}
\usepackage[english]{babel}
\usepackage{wrapfig}
\usepackage{animate}
\usepackage{xcolor}
\usepackage{booktabs}  
\usepackage{array}     
\usepackage{times}
\usepackage[T1]{fontenc}
\usepackage{color, colortbl}
\usepackage{tikz}
\usepackage[usenames, dvipsnames]{xcolor}
\usetikzlibrary{arrows,shapes}
\usepackage{url}
\usepackage{hyphenat}
\usepackage{makecell}
\usepackage[Conny]{fncychap}
\usepackage{lipsum}
\usepackage{mathptmx}
\usepackage{mathtools}
\usepackage[T1]{fontenc}
\usepackage[%
  colorlinks=true,
  urlcolor=blue,
  linkcolor=blue,
  citecolor=blue
]{hyperref}
\DeclareUnicodeCharacter{2212}{-}
\newcommand{\orcid}[1]{\href{https://orcid.org/#1}{\resizebox{10px}{!}{\includegraphics{orcid.png}}}}

\begin{document}
\title{A Casimir-like probe for 4D Einstein-Gauss-Bonnet gravity}

\author{Syed Masood$^{1,2,3}$} 
\email{quantummind137@gmail.com}
\affiliation{$^{1}$Zhejiang University/University of Illinois at Urbana-Champaign Institute (the ZJU-UIUC Institute),
Zhejiang University, $718$ East Haizhou Road, Haining $314400$, China,}
\affiliation{$^{2}$Department of Physics, School of Science and Research Center for Industries of the Future, Westlake University, Hangzhou $310030$, P. R. China,}
 \affiliation{$^{3}$Institute of Natural Sciences, Westlake Institute for Advanced Study, Hangzhou $310024$, P. R. China.}
\begin{abstract}
 Virtual transitions in a Casimir-like configuration are utilized to probe quantum aspects of  four-dimensional Einstein-Gauss-Bonnet (4D EGB) gravity. This study employs a quantum optics-based approach, wherein an Unruh-DeWitt detector (modeled as a two-level atom) follows a radial timelike geodesic, falling freely into an uncharged, nonrotating black hole described by 4D EGB gravity, becoming thermalized in the usual Unruh manner. The black hole, asymptotically Minkowskian, is enclosed by a Casimir boundary proximate to its horizon, serving as a source for accelerated field modes that interact with the infalling detector. Observations are conducted by an asymptotic infinity observer, assuming a Boulware field state.
Our numerical analysis reveals that, unlike in Einstein gravity, black holes in 4D EGB gravity can either enhance or suppress the intensity of acceleration radiation, contingent upon the Gauss-Bonnet coupling parameter $\alpha$. Specifically, we observe radiation enhancement for negative $\alpha$ and suppression for positive $\alpha$. These findings offer substantial insights into quantifying the influence of higher-curvature contributions on the behavior of quantum fields in black hole geometries within a 4D spacetime.
    \end{abstract}
\date{\today}
\maketitle

\section{Introduction}

Considerable efforts have been made over the past few decades to uncover the deep connection between quantum mechanics, gravity, and thermodynamics \cite{Birrell:1982ix,2009qftc.book.....P}. Among these endeavors, the discovery of Hawking radiation from black holes \cite{1975CMaPh..43..199H} and the Unruh effect for accelerated observers in flat Minkowski spacetime \cite{Unruh:1976db,RevModPhys.80.787} stand out as pivotal. Another significant phenomenon is Parker's idea of particle emission due to the expansion of the Universe \cite{2009qftc.book.....P}. In all these cases, the quantum state of the field is altered by a dynamic background spacetime geometry or the state of motion, resulting in the creation of real particles---an effect arising from the violation of Poincaré invariance \cite{Birrell:1982ix}. This is similar to the dynamical Casimir effect (DCE) \cite{1970JMP....11.2679M,Dodonov:2020eto}, where accelerated plates or boundaries induce the quantum vacuum to radiate particles. Consequently, this scenario fosters a rich intersection of quantum fields, boundaries, and spacetime geometries \cite{Scully:2017utk,Svidzinsky:2018jkp,Lopp:2018lxl,MasoodASBukhari:2023flr}.

\indent With the advent of precise experimental and observational setups, it has become possible over the decades to test Einstein's general relativity (GR) in extreme gravity regimes. So far, GR has consistently matched observational data, with milestone achievements including gravitational wave detection \cite{LIGOScientific:2016aoc,LIGOScientific:2016lio}, black hole shadows \cite{EventHorizonTelescope:2019dse}, and neutron star mergers \cite{Sanger:2024axs}. However, physicists have long recognized that GR cannot address certain fundamental issues in the Universe, such as the existence of singularities, cosmological acceleration, dark matter, and a consistent merger of quantum mechanics and gravity. Thus, it is evident that a framework beyond GR is needed to resolve these challenges \cite{Capozziello:2011et, Berti:2015itd, Shankaranarayanan:2022wbx}.\\
\indent 
Several alternatives to GR predict additional higher-curvature contributions to the gravitational action. A significant framework within this class of models originates from the works of Lanczos \cite{Lanczos:1938sf} and Lovelock \cite{Lovelock:1971yv, Lovelock:1972vz}, leading to the well-known Einstein-Gauss-Bonnet (EGB) theory. It has been established that EGB gravity does not introduce modifications to gravitational dynamics unless coupled with additional field degrees of freedom or in spacetime dimensions D$\geq 5$. One example of such additional fields is the dilaton field \cite{Blazquez-Salcedo:2016enn, Konoplya:2019fpy, Maselli:2014fca, Ayzenberg:2013wua}.
In addition to this, EGB gavity theories yield equations of motion that are quadratic in metric tensor. This quadratic nature  is a unique feature  of EGB gravity among all other alternatives to GR. The interesting coincidence is  that the low energy effective descriptions of heterotic string theories also posit quadratic contributions to the dynamics of Einstein gravity \cite{Zwiebach:1985uq,Gross:1986iv,Gross:1986mw}.   It may be noted that the quadratic nature of equations of motion suffice to get rid of Ostrogradsky instability \cite{Ostrogradsky:1850fid} and thus guarantees physicality of the dynamics.  
Furthermore, EGB gravity theories are characterized by equations of motion that are quadratic in the metric tensor. This quadratic nature distinguishes EGB gravity from other alternatives to GR. An intriguing coincidence arises in that the low-energy effective descriptions of heterotic string theories also incorporate quadratic contributions to the dynamics of Einstein gravity \cite{Zwiebach:1985uq,Gross:1986iv,Gross:1986mw}. Importantly, the quadratic form of the equations of motion resolves the Ostrogradsky instability \cite{Ostrogradsky:1850fid}, ensuring the physical viability of the theory.\\
\indent 
Recently, Glavan and Lin \cite{Glavan:2019inb} addressed the question of Gauss-Bonnet (GB) contributions in 4-dimensional spacetime geometry by proposing a specific rescaling of the GB coupling parameter $\alpha \rightarrow \alpha/\left(\text{D}-4\right)$, where $D$ denotes the spacetime dimensionality. This rescaling ensures a well-defined limit as $D \rightarrow 4$. The resulting model maintains quadratic behavior to prevent Ostrogradsky instability, yet it departs from the implications of the well-known Lovelock theorem \cite{Lovelock:1971yv, Lovelock:1972vz, Lanczos:1938sf}. It is noteworthy that no additional field coupling is required in this model. As a new phenomenological competitor to Einstein's General Relativity (GR), this model has sparked rigorous debates over the years. Some investigations include consistency checks \cite{Lu:2020iav, Hennigar:2020lsl, Gurses:2020ofy}, studies of black hole shadows and quasinormal modes \cite{Konoplya:2020bxa, Kumar:2020owy, Wei:2020ght}, analysis of geodesics \cite{Guo:2020zmf}, particle accelerator models \cite{NaveenaKumara:2020kpz}, and a wide array of thermodynamic analyses \cite{Wei:2020poh, HosseiniMansoori:2020yfj, EslamPanah:2020hoj, Hegde:2020xlv, Hegde:2020yrd, Kumar:2024bls, Kumar:2023ijg, Ghosh:2020tgy, Masood:2024yio}. A comprehensive overview of 4D-EGB gravity, covering its various aspects, can be found in a review article by Fernandes \textit{et al.} \cite{Fernandes:2022zrq}.\\
\indent 
Recognizing the significance of the findings in Ref. \cite{Glavan:2019inb}, we are driven to investigate the potential quantum radiative signatures of 4D EGB gravity using elements from quantum optics and Casimir physics. Our approach involves a quantum optical cavity positioned with one end near a black hole horizon and the other at asymptotic infinity. Within this setup, a two-level Unruh-DeWitt detector (an atom) falls freely towards the black hole. Virtual transitions arising from the interaction between the detector and the field lead to acceleration radiation, which carries distinct imprints of the underlying gravitational background. Such a setup has been discussed in Ref. \cite{Scully:2017utk}, where it was demonstrated that, under appropriate initial conditions, a detector near a Schwarzschild black hole emits radiation with a thermal spectrum. This unique radiative emission, known as Horizon Brightened Acceleration Radiation (HBAR), occurs when the detector is in free fall towards the black hole. This concept has been further explored in various contexts, revealing profound connections between the equivalence principle, quantum optics, and the Hawking-Unruh effect \cite{Scully:2017utk, Fulling:2018lez, Ben-Benjamin:2019opz, Chatterjee:2021fue, Sen:2022tru}. It also underscores connections to the Dynamical Casimir Effect (DCE) and moving mirror models \cite{Anderson:2015iga, Good:2015jwa}, frequently employed in studying quantum field behavior in curved spacetimes. 
But while the original work in Ref. \cite{Scully:2017utk} considers detectors moving along timelike geodesics, subsequent studies have shown that similar phenomena can occur for detectors following null geodesics \cite{Chakraborty:2019ltu}. This novel radiative emission phenomenon can be attributed to the near-horizon physics and conformal quantum mechanics of black holes \cite{Camblong:2020pme, Azizi:2020gff, Azizi:2021qcu, Azizi:2021yto, Sen:2022cdx, Sen:2023zfq}. \\
\indent 
Given that quantum field dynamics can elucidate the nature of underlying spacetime geometry \cite{Birrell:1982ix, 2009qftc.book.....P}, we view the aforementioned setup as a potential avenue to probe 4D EGB gravity at a deeper level. Through numerical analysis, we demonstrate that 4D EGB gravity can imprint distinct features on the radiation spectrum compared to Einstein's GR, encompassing both negative and positive values of the Gauss-Bonnet coupling parameter.\\
\indent
The structure of the paper is as follows. The next Sec. \ref{geom} introduces the basics of 4D EGB black hole geometry, accompanied by discussions on the wave equation and the vacuum field state. In Sec. \ref{secPEGB}, we compute the excitation probability or the detector response function of the falling detector. Sec. \ref{discussionEGB} explores possible interpretations of our numerical findings. Finally, conclusions are drawn in Sec. \ref{secconclusions}.

\textcolor{black}{
\section{Horizon Brightened acceleration radiation: An outlook}
The core idea behind  radiation emission from the vacuum in the typical Hawking-Unruh paradigm  is the fact that vacuum fluctuations do not follow Poincar\'e invariance in presence of external influences, such as gravity or acceleration \cite{Birrell:1982ix}. Hence it is tantamount to say that quantum vacuum gets polarized due to these factors.  Beyond famous Hawking-Unruh radiation, this produces myriad of  non-stationary QED effects \cite{MasoodASBukhari:2023flr,Nation:2011dka}, including  dynamical Casimir effect as mentioned in  Introduction. Out of many interesting mechanisms, one  possible idea  is to utilize  moving boundaries. In principle, a moving boundary (or mirror)  potentially alters the
structure of vacuum, resulting  in the creation
and annihilation of radiation quanta \cite{Haro:2006zz}. These models are very crucial for understanding particle production in many cosmological scenarios and radiation emitted from black holes\cite{MasoodASBukhari:2023flr}. Our investigation in the present work revolving around Ref.\cite{Scully:2017utk} is grounded in the concept of these moving mirrors or accelerated boundaries.
\begin{figure}[tbhp]
\centering
\includegraphics[width=\linewidth, height=8.6cm]{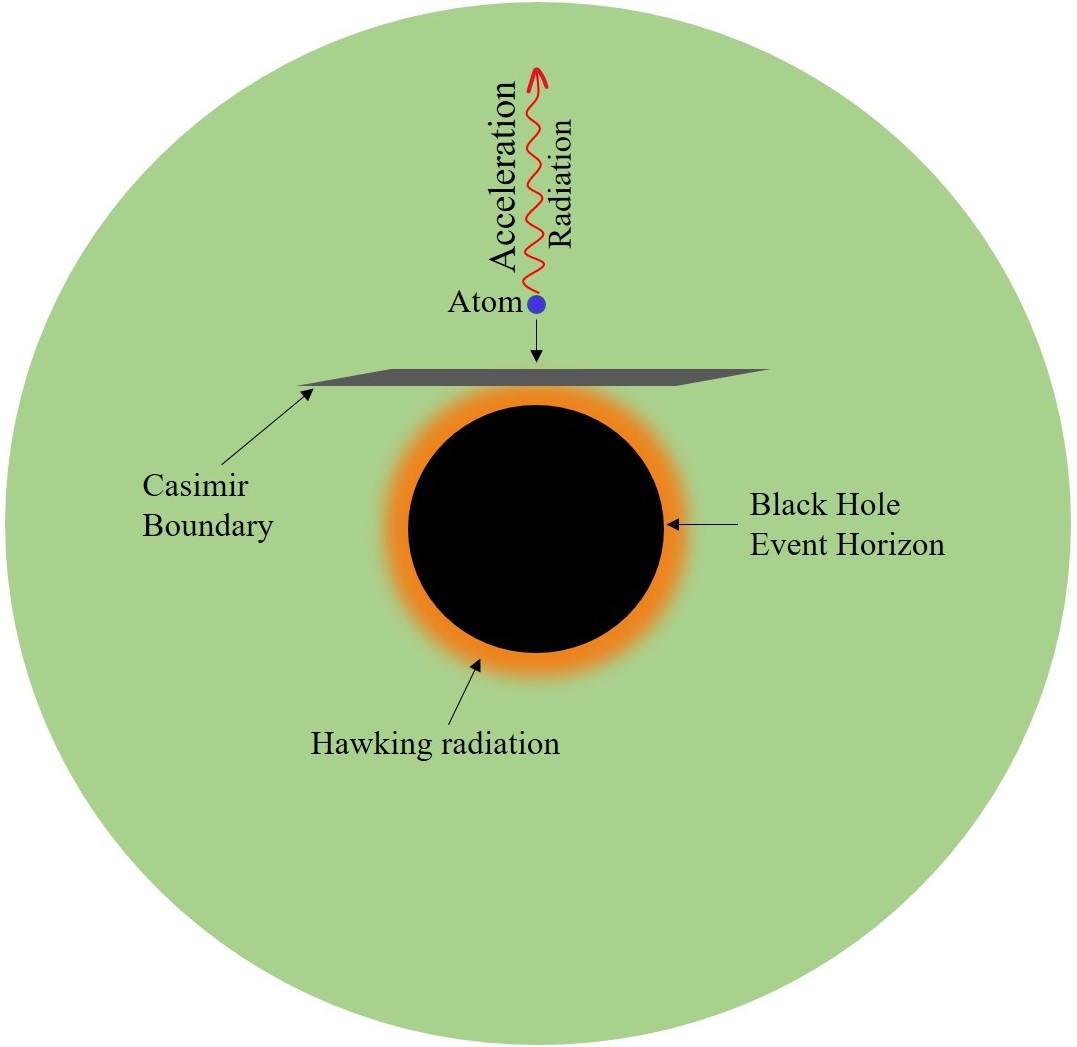}
\caption{\textcolor{black}{Schematic view  of acceleration radiation emitted from a two-level atom falling into a black hole. The Casimir boundary provides accelerated field modes with which the atom interacts with, besides shielding it from Hawking radiation. Hence, no quanta in the accelerated radiation has any Hawking contribution.}}
\label{schematic}%
\end{figure}
\subsection{The Model}
\indent To best comprehend the workings of our model and subsequent investigations, we first provide a pictorial representation in Fig. \ref{schematic} to put things as easy as possible. With that, we believe it captures the essence of the setup introduced in \cite{Scully:2017utk}.\\
\indent The configuration assumes a black hole with its event horizon as indicated in Fig.\ref{schematic} above which hovers one end  (mirror or Casimir boundary) of a cavity\footnote{\textcolor{black}{Consider the mirror being held in position firmly against black hole's gravity by some rockets!}} whose other end is assumed to be fixed somewhere at asymptotic infinity. A two-level atom acting as a typical Unruh-De Witt detector falls freely from infinity toward the event horizon of the black hole.  Thus, in accordance with the equivalence principle of general relativity, atom is in an inertial frame of reference (zero four-acceleration), while the Casimir boundary is in an accelerated frame of reference. An observer, located at infinity,  defines the field around the black hole in its vacuum state. Specifically, as such observer is at asymptotically flat region, the appropriate vacuum  for such an observer is the Boulware state \cite{Birrell:1982ix}. Since atom is in inertial frame, it is not expected to emit any radiation, and this the essence of quantum field theory in curved spaces \cite{Birrell:1982ix}. However, as argued in \cite{Scully:2017utk}, the vacuum field modes interacting with the mirror are polarized in a similar manner to that of Hawking radiation. Thus there is a relative acceleration between atom and the Casimir boundary which gives rise to the acceleration radiation detected by the asymptotic observer. The Casimir boundary, in addition to providing accelerating field modes, also shields the atom from any Hawking quanta (shown as orange colored ring). The boundary can be envisioned by   either assuming a reflecting mirror as shown in Fig.\ref{schematic} or an enclosing spherical surface shrouding the entire  black hole \cite{Bukhari:2022wyx,Bukhari:2023yuy}.}

\textcolor{black}{
\subsection{The Significance}
It is quite intriguing to see from above that normally, such observers with Boulware vacuum state do not observe any Hawking quanta. This radiation emission  thus has purely a quantum optical origin and has surprising similarities to that of Hawking radiation. The radiation has a thermal Planck distribution with a  characteristic temperature and gives rise to an entropy-area correspondence. While defining the associated entropy, it becomes crucial to consider area lost by the black hole due to the  radiation. From the relation \cite{Scully:2017utk},  
\begin{eqnarray}
    \dot{S}_{\rm HBAR }=\frac{k_{\rm B}c^3}{4\hbar G}\dot{A},
\end{eqnarray}
where $k_{\rm B}, c, \hbar$ and $G$ represent Boltzmann constant, velocity of light, Planck's constant Newtons' gravitational constant, respectively, one can see the rate of change of the entropy being related to the rate of loss of area $A$ of the black hole through HBAR flux. Despite these similarities  between Hawking radiation and HBAR, certain parameters like phase correlations between quanta make HBAR different from Hawking emission, avoiding any information loss paradox.\\
\indent  Above setup undoubtedly  provides new insights into Hawking-Unruh effect and equivalence principle by assuming some kind of atom-mirror symmetry \cite{Fulling:2018lez,Svidzinsky:2018jkp}. However, certain extreme scenarios in quantum gravity, where the field satisfies modified dispersion relation \cite{Kempf:1994su}, might indicate limited regimes of applicability of HBAR emission \cite{Chatterjee:2021fue}.   }
\textcolor{black}{\section{Our spacetime geometry and the choice of field modes}}\label{geom}

\textcolor{black}{In Einstein gravity, the 4D Einstein-Hilbert action reads
   \begin{eqnarray}
       S_{\rm EH}=\int \mathrm{d}^4x\sqrt{-g}R,
   \end{eqnarray}
   where $R$ denotes Ricci curvature, and $g$ is the determinant of metric tensor $g_{\mu\nu}$. According to  Lovelock theorem \cite{Lovelock:1971yv, Lovelock:1972vz, Lanczos:1938sf}, GR is the unique gravitational theory  in four dimensions because it satisfies the criteria of being diffeomorphism-invariant, possesses metricity, and its equations of motion are all  second-order. In higher spacetime dimensions, GB action
   \begin{eqnarray}\label{actionGB}
       S_{\rm GB}=\int \mathrm{d}^{D}x\sqrt{-g}\left(R+\alpha \mathcal{G}\right)
   \end{eqnarray}
   satisfies these conditions also. Here, $ \mathcal{G}=R^2-4R_{\mu\nu}R^{\mu\nu}+R_{\mu\nu\rho\sigma}R^{\mu\nu\rho\sigma}$ is the GB invariant term, $R^{\mu\nu}$ and $R^{\mu\nu\rho\sigma}$ denote the Ricci and Riemann tensors, respectively. If one varies Eq. (\ref{actionGB}) with respect to $g_{\mu\nu}$, one gets gravitational field equations
   \begin{eqnarray}\label{FEGB}
      R_{\mu\nu}-\frac{1}{2}Rg_{\mu\nu}+\alpha H_{\mu\nu}=T_{\mu\nu},
   \end{eqnarray}
   where we wrote 
   \begin{align}\label{FEHMN}
H_{\mu\nu}=&2RR^{\mu\nu}-4R_{\mu\sigma}R^{\sigma}_{\nu}-4R_{\mu\sigma\nu\rho}R^{\sigma\rho}\\ 
       &-2R_{\mu\sigma\nu\delta}R^{\sigma\rho\delta}_{\nu}-\frac{1}{2}\mathcal{G}g_{\mu\nu},
  \end{align}
   and $T_{\mu\nu}$ represents the energy-momentum tensor. }\\
\indent The static, spherically symmetric metric of an uncharged and nonrotating  black hole in 4D EGB gravity \textcolor{black}{gotten from Eq. \ref{FEGB}} is given by \cite{Fernandes:2022zrq} 
\begin{eqnarray}
ds^{2}=-f(r) \mathrm{d}t^{2}+\frac{1}{f(r)} \mathrm{d}r^{2}+r^{2}( \mathrm{d}\theta ^{2}+\sin^{2}\theta  \mathrm{d}\phi ^{2}),
\label{metricDM}
\end{eqnarray}
where\footnote{\textcolor{black}{We use natural units $c=G=\hbar=k_{\rm B}=1$ throughout from now on.}}
\begin{eqnarray}\label{frGB}
f(r)=1+\frac{r^2}{2\alpha}\left(1\pm\sqrt{1+\frac{8\alpha M}{r^3}}\right),\ \ \ 
\end{eqnarray}
where $\pm$ sign inside brackets denotes Gauss-Bonnet (GB) and GR branches, respectively. Here, we focus solely on the GR branch, as the GB branch is deemed unphysical \cite{Fernandes:2022zrq}. To determine the event horizon radius, we set
\begin{eqnarray}\label{bound}
    f(r)=1+\frac{r^2}{2\alpha}\left(1-\sqrt{1+\frac{8\alpha M}{r^3}}\right)=0,
\end{eqnarray}
which yields 
\begin{eqnarray}\label{rad}
    r_{\pm}=M\pm \sqrt{M^2-\alpha},
\end{eqnarray}
\textcolor{black}{of which the one with plus sign is the (outer) horizon of the black hole. The parameter $\alpha$ can take both positive and negative values, as indicated in Refs. \cite{Guo:2020zmf, Konoplya:2020bxa} (also see \cite{Glavan:2019inb}). Moreover, for $\alpha>0$, there is a minimum mass of the black hole given by
\begin{eqnarray}
    M_{\rm min}=\sqrt{\alpha},
\end{eqnarray}
below which $f(r)$ does not vanish and one encounters naked singularities. The negative values of $\alpha$ have been shown to be tightly constrained \cite{Charmousis:2021npl,Clifton:2020xhc}. In this paper, working in natural units, we take range of $\alpha$ as $[-1,1]$, consistent with Eq.\ref{rad}. If one likes to compute parameters in dimensional units, then one notes that mass $M$ in Eq. \ref{bound} has to be replaced by $GM/c^2$, which in view of Eq. \ref{rad} necessitates $\alpha<G^2M^2/c^4$ to avoid any naked singularity. The bounds for  values of $\alpha$ as noted in \cite{Charmousis:2021npl} for the case of black holes vis-à-vis gravitational wave event GW$190814$ \cite{LIGOScientific:2020zkf} are $-10^{-30}\rm m^2\leq \alpha< 5.9\times 10^{7}\rm m^2$, which are tighter (especially for negative $\alpha$) than the bounds provided in  Ref.\cite{Clifton:2020xhc} with $-10^{10}<\alpha<10^9\rm m^2$.  The range $\alpha\in [-1,1]$ we consider in this work is in natural units which makes them arbitrary and hence suitable for any constraints and system of units. However,  all calculations done in this paper should be viewed in compliance with the bounds  provided in Refs. \cite{Charmousis:2021npl,Clifton:2020xhc}. }

\begin{figure*}[t]
\centering
\includegraphics[width=.85\linewidth, height=7.cm]{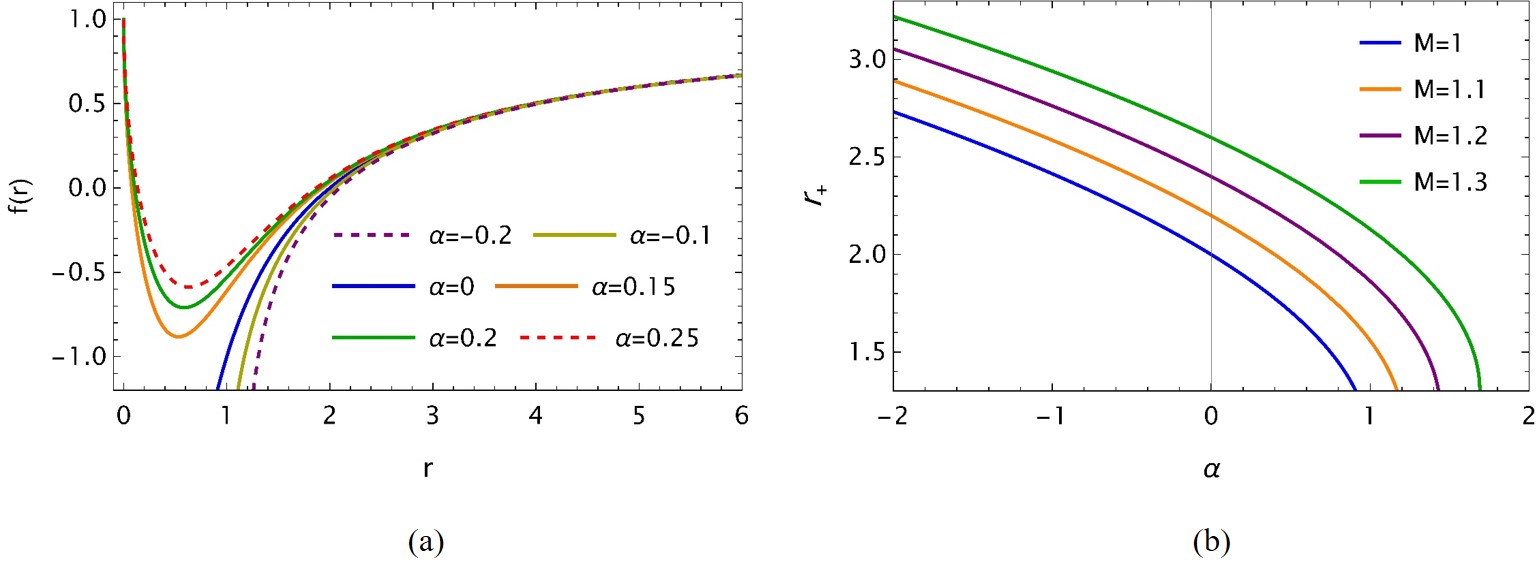}
\caption{Impact of GB coupling $\alpha$ on (a) $f(r)$ and (b) horizon radius $r_{+}$ to show distinct nature of $\alpha$ for its positive and negative values.}%
\label{frrgplot}%
\end{figure*}

It is evident that a positive GB coupling constant $\alpha$ decreases the black hole horizon radius, whereas a negative $\alpha$ increases it. The limit $\alpha = 0$ corresponds to the Schwarzschild black hole in GR. These relationships are illustrated graphically in Fig. \ref{frrgplot}.

We also note that $ r_{+}r_{-} = \alpha $, and as $ r \rightarrow 0 $, the metric components remain finite. This can be observed from Eq. (\ref{frGB}), where $ \lim_{r \to 0} f(r) = 1 $. However, the finiteness of the metric components does not guarantee the absence of singularities due to the fact that the Ricci scalar $ R $ and the Kretschmann scalar $ R_{\mu\nu\sigma\delta}R^{\mu\nu\sigma\delta} $ vary as $ R \propto r^{-3/2} $ and $ R_{\mu\nu\sigma\delta}R^{\mu\nu\sigma\delta} \propto r^{-3} $, respectively. It should be noted that for the Schwarzschild case, the Kretschmann scalar near $ r=0 $ varies as $ r^{-6} $, indicating that the GB contribution significantly weakens the singularity by several orders of magnitude \cite{Fernandes:2022zrq}.\\
\indent \textcolor{black}{Although $4$D EGB gravity has many interesting phenomenological predictions compared to GR, its overall structure and consistency has been point of major debate among researchers. For example, the tensor $H_{\rm \mu\nu}$ given by Eq.(\ref{FEHMN}) has been shown not to possess an inherently $4$D
description in terms of a covariantly-conserved rank-two tensor in $4$D \cite{Gurses:2020ofy,Gurses:2020rxb,Arrechea:2020evj,Arrechea:2020gjw}. Moreover, the possibility of a nontrivial D$\rightarrow 4$ limit of Eq.\ref{actionGB} has also been investigated using tree-level graviton scattering amplitudes \cite{Bonifacio:2020vbk}. It has been found that $4$D EGB gravity in some  sense  can be understood as a description arising out  of a scalar field, hence making the theory a part of scalar-tensor theories \cite{Bonifacio:2020vbk,Fernandes:2020nbq}. Furthermore, the divergences in the on-shell action   \cite{Mahapatra:2020rds} makes it impossible to variationallly complete field equations \cite{Hohmann:2020cor}. In view of these concerns and criticisms, the regularization of $4$D EGB has been a subject of intense investigation. For a detailed outlook, we refer the interested reader to the Refs.\cite{Fernandes:2020nbq,Fernandes:2021dsb,Hennigar:2020lsl,Kobayashi:2020wqy,Lu:2020iav,Aoki:2020lig}.}

\subsection{Detector trajectories}

In this section, we analyze the geodesics of the detector to compute both the coordinate time and proper (conformal) time that describe the timelike trajectory of the infalling (massive) detector. Generally, for a given Christoffel connection $\Gamma_{\rho\sigma}^{\mu}$, the complete geodesic equations are expressed as \cite{1992mtbh.book.....C}
\begin{eqnarray}
 \frac{ \mathrm{d}^2 x^{\mu}}{ \mathrm{d}\tau^2}+\Gamma_{\rho \sigma}^{\mu}\frac{ \mathrm{d}x^{\rho}}{\mathrm{d}\tau}\frac{ \mathrm{d}x^{\sigma}}{ \mathrm{d}\tau}=0.
\end{eqnarray}
Our spacetime geometry of interest exhibits spherical symmetry, and we restrict our analysis to the radial motion of the detector in the equatorial plane. Therefore, we set $\theta = \pi/2$, which implies $\dot{\theta} = 0$ and $\dot{\phi} = 0$. Consequently, the following conservation equations hold:
\begin{eqnarray}
 \Big(\frac{ \mathrm{d}r}{ \mathrm{d}\tau}\Big)^2=\mathcal{E}^2-f(r),\ \  \Big(\frac{ \mathrm{d}r}{ \mathrm{d}t}\Big)^2=\left[\frac{f(r)}{\mathcal{E}}\right]^2\left[\mathcal{E}^2-f(r)\right].
\end{eqnarray}
Note that $\mathcal{E}$ is a constant representing the specific energy of the detector. It is determined by the initial boundary conditions of the geodesic motion, given by $\mathcal{E}^2 = f(r) \big|_{\text{max}}$. Since we assume that the detector started its motion from asymptotic infinity, where the spacetime is asymptotically Minkowski flat ($r \rightarrow \infty$ implies $f(r) \big|_{\text{max}} = 1$), these constraints from the above equations lead to
\begin{eqnarray}\label{midt}
 \Big(\frac{ \mathrm{d}r}{ \mathrm{d}\tau}\Big)^2=1-f(r),\ \ \Big(\frac{ \mathrm{d}r}{ \mathrm{d}t}\Big)^2=f^2(r)\left[1-f(r)\right].
\end{eqnarray}
It should be emphasized that $\mathcal{E}$, which is related to the maximum of $f(r)$, is the same for both GR and 4D EGB gravity. This value of $\mathcal{E}$ corresponds to asymptotic infinity, where both GR and 4D EGB theories reproduce flat Minkowski geometry.

\begin{figure*}[t]
\centering
\includegraphics[width=\linewidth, height=14cm]{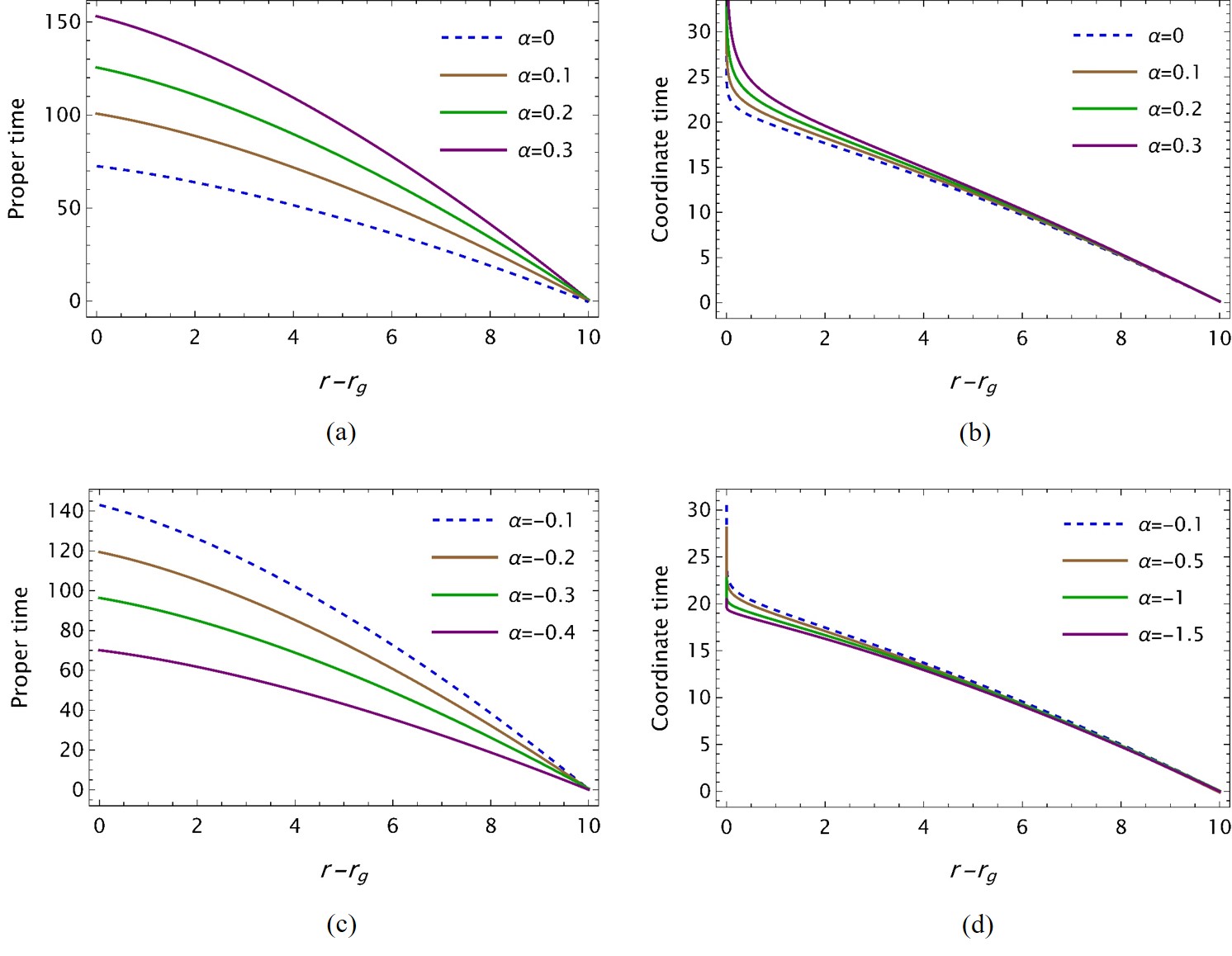}
\caption{\textcolor{black}{The radial dependence of coordinate time $ t $ and proper time $ \tau $ illustrates the influence of the Gauss-Bonnet coupling parameter $\alpha$. Panels (a) and (c) depict the proper time for positive and negative $\alpha$, respectively. Similarly, panels (b) and (d) show the coordinate time for positive and negative $\alpha$, respectively. For all plots, we set black hole mass $M=1$. }}%
\label{time}%
\end{figure*}

Now, integrating Eq. (\ref{midt}) along the radial trajectories from some arbitrary initial point $r_{\rm i}$ to a final point $r_{\rm f}$ (where $r_{\rm i} > r_{\rm f}$), we obtain
\begin{eqnarray}\label{t}
 \tau =- \int_{r_{\rm i}}^{r_{\rm f}} \frac{ \mathrm{d}r}{\sqrt{1-f(r)}},\ \ t=-\int_{r_{\rm i}}^{r_{\rm f}}\frac{ \mathrm{d}r}{f(r)\sqrt{1-f(r)}}.
\end{eqnarray}
We now substitute Eq. (\ref{frGB}) into Eq. (\ref{t}) in order to compute $\tau$, resulting in
\begin{align}
  \tau=\frac{2 r \sqrt{\sqrt{1+\frac{8 \alpha M}{r^3}}-1} \tan ^{-1}\left(\frac{\sqrt{\sqrt{1+\frac{8 \alpha M}{r^3}}-1}}{\sqrt{2}}\right)-2 \sqrt{2} r}{3 \sqrt{\frac{r^2
   \left(\sqrt{1+\frac{8 \alpha M}{r^3}}-1\right)}{\alpha}}}+\tau_{0}.
\end{align}
Here, $\tau_{0}$ serves as an integration constant, the insignificance of which we establish for the final detector response, as detailed in Sec. \ref{secPEGB}. However, the complexity of the integral for $ t $ precludes straightforward analytical computation. Consequently, we resort to numerical methods and present the outcomes in Sec. \ref{secPEGB}. Fig. \ref{time} illustrates the plots of $ \tau $ and $ t $.

The plots clearly illustrate that $ t $ and $ \tau $ exhibit typical Schwarzschild-like behavior. Specifically, $ t $, which represents the time measured by an asymptotic observer, diverges as the detector approaches the black hole horizon, located at zero on the rescaled radial coordinate $ r - r_{+} $. This divergence signifies that, from the perspective of this observer, the detector never actually crosses the horizon. 
In contrast, $ \tau $ remains finite at the horizon $ r - r_{+} $, indicating that from the detector's own frame of reference, it crosses the horizon in a finite amount of proper time. This disparity highlights the causal structure of black hole horizons and is recognized as gravitational time dilation.

Furthermore, in 4D EGB gravity, the coupling parameter $\alpha$ influences the behavior of $ t $ and $ \tau $. For positive $\alpha$, which reduces the black hole size as discussed in Sec. \ref{geom}, it takes longer for the detector to approach the horizon as $\alpha$ increases. Conversely, for negative $\alpha$, which inflates the black hole size, the situation is reversed.

\subsection{Defining the vacuum state}

The response function, or excitation probability, to be calculated in Sec. \ref{secPEGB}, quantifies the detector-field coupling. To achieve this, we must obtain the appropriate field mode by solving the wave equation on the specified spacetime background. Here, we consider the simplest test field: a massless spin-0 Klein-Gordon field, minimally coupled to the spacetime geometry, described by $\nabla_{\mu}\nabla^{\mu} \Phi = 0$ \cite{Birrell:1982ix}.
 Given the spherical symmetry of the spacetime and the presence of a timelike Killing vector $\partial_{t}$, we have 
 $\Phi=\frac{1}{r}Y_{l}(\theta,\phi)\psi(t,r)$,
with $ Y_{l} $ denoting spherical harmonics and $ l $ representing the multipole number. The radial part of the solution, after neglecting the angular dependence $(l=0)$, satisfies the following Schrödinger-like wave equation
\begin{eqnarray}\label{WE}
 \Big(-\frac{\partial^2 }{\partial t^2}+\frac{\partial^2 }{\partial r_{*}^2}\Big)\psi(t,r)=V(r)\psi(t,r).
\end{eqnarray}
Here, $ r_{*} $ denotes the Regge-Wheeler tortoise coordinate, a useful parameter for describing the propagation of test fields in black hole geometries, defined by \cite{1992mtbh.book.....C}
\begin{align}
    r_{*}&=\int \frac{ \mathrm{d}r}{f(r)}\\
    \label{rstar}
    &=\int \frac{ \mathrm{d}r}{1+\frac{r^2}{2\alpha}\left(1-\sqrt{1+\frac{8\alpha M}{r^3}}\right)},
\end{align} 
where we utilized Eq. (\ref{frGB}). Additionally, $ V(r) $ represents the effective potential experienced by the field, often describing scattering effects in black hole spacetimes \cite{Futterman:1988ni}. However, given our focus on the simplest scenario possible, as also demonstrated in Refs. \cite{Scully:2017utk,Bukhari:2022wyx}, $ V(r) $ can be neglected. One approach to achieve this is by assuming that the frequency $ \nu $ of the field mode is sufficiently large, enabling it to overcome the potential barrier imposed by the spacetime. Consequently, the field mode simplifies to
\begin{eqnarray}\label{fieldmodeDM}
 \psi(t,r)=\exp{\left[i\nu(t- r_{*})\right]}.
\end{eqnarray}
This represents a normalized outgoing field mode with frequency $ \nu $, as observed by an asymptotic infinity observer, qualifying as a Boulware field state. The ingoing field modes generated propagate towards the boundary at the black hole horizon and are lost.\\
 \indent 
The Boulware field mode described above is an approximate field state obtained by neglecting $ V(r) $ and assuming $ \nu $ to be very large. This assumption serves as one of the initial conditions required for the existence of HBAR emission \cite{Scully:2017utk}. Generally, in the context of black holes, multiple vacuum states are utilized due to the absence of a unique vacuum state in curved spacetime. This leads to various notions of vacuum states, such as the Unruh vacuum, Hartle-Hawking vacuum \cite{Birrell:1982ix,2009qftc.book.....P}, and others.
In principle, there should be an infinite number of possible vacuum states due to the violation of Poincaré invariance in curved spacetimes \cite{Birrell:1982ix}. In contrast, for Minkowski space, where the field satisfies Poincaré invariance, the vacuum state remains  same for all inertial observers. In our scenario, the choice of the Boulware vacuum state arises because the observations are made by an asymptotic observer, for whom the Boulware field state is most appropriate. In this context, no Hawking radiation is detected by the observer. Moreover, the black hole is assumed to be entirely enclosed by a Casimir boundary, which effectively prevents any potential Hawking quanta from mixing with  HBAR flux \cite{Scully:2017utk}. This distinction ensures that HBAR emission is fundamentally different from Hawking radiation.\\
\indent 
Additionally, we have excluded $ l=0 $ modes for simplicity. However, considering a smaller $ \nu $ such that $ V(r) \neq 0 $ would lead to the emergence of scattering effects, potentially necessitating the inclusion of greybody factors \cite{Sakalli:2022xrb}. Nevertheless, we argue that such inclusions would lead to the deviation from the primal essence of HBAR emission, which occurs under specific boundary conditions as emphasized in Refs. \cite{Scully:2017utk,Ben-Benjamin:2019opz}.

\section{Detector Response }\label{secPEGB}

As discussed in the preceding section, the field is in a Boulware vacuum state, ensuring that no Hawking radiation is observed by the asymptotic observer. By neglecting the angular dependence of the field modes, the detector-field interaction Hamiltonian can be expressed as follows \cite{Scully:2017utk}:
\begin{align}\label{hamiltonianEGB}
 \hat{H}(\tau)&=\hbar g \big[\hat{a}_{\nu}\psi\left[t(\tau),r(\tau)\right]+\mathrm{H.C.}\big]\big[\hat{\sigma}(\tau)e^{-i\omega\tau}+\mathrm{H.C.}\big].
\end{align}
Here, $\hat{a}_{\nu}$ is the annihilation operator for the field modes, $\hat{\sigma}$ is the detector lowering operator, and $\mathrm{H.C.}$ denotes the Hermitian conjugate. Here, $ g $ is a detector-field coupling parameter indicating the strength of the interaction and can be taken as a constant for a massless Klein-Gordon field (spin-0). 

Assuming that the detector is initially in the ground state $ |b\rangle $, the probability that it transitions to an excited state $ |a\rangle $ with the emission of a field quantum of frequency $ \nu $ is given by
\begin{eqnarray}\label{exc1EGB}
 \Gamma_{\rm exc}=\frac{1}{\hbar^2}\bigg|\int  \mathrm{d}\tau\, \langle 1_{\nu},a|H(\tau)|0,b\rangle\bigg|^2.
\end{eqnarray}
Utilizing time-dependent perturbation theory, such a process is typically prohibited in quantum optics due to energy conservation principles. However, in non-inertial frames influenced by acceleration and gravity, these virtual processes can occur owing to counter-rotating terms in the Hamiltonian \cite{Ben-Benjamin:2019opz}, as exemplified by the Unruh effect \cite{Unruh:1976db}.
By employing Eq. (\ref{hamiltonianEGB}) and performing some additional straightforward computations, Eq. (\ref{exc1EGB}) can be reexpressed as
\begin{align}\nonumber
 \Gamma_{\rm exc}&=g^2\bigg|\int  \mathrm{d}\tau\,  \psi^{*}(t(\tau),r(\tau))e^{i\omega\tau}\bigg|^2\\
 \label{pexx1}
 &=g^2\bigg|\int  \mathrm{d}r \bigg(\frac{ \mathrm{d}\tau\, }{ \mathrm{d}r}\bigg) \psi^{*}(r)e^{i\omega\tau}\bigg|^2.
  \end{align}
Simplifying further, we arrive at 
\begin{widetext}
  \begin{align}\label{probalpha}
  \Gamma_{\rm exc}&= g^2\left|\int\limits_{\infty}^{r_{+}}  \mathrm{d}r\,  \exp{\left[i\nu\left\{t(r)- r_{*}(r)\right\}\right]} 
    \frac{1}{\sqrt{\frac{r^2}{2\alpha}\left( \sqrt{1+\frac{8\alpha M}{r^3}}-1\right)}}
    \exp{\left[i\omega\left\{\frac{2 r \sqrt{\sqrt{1+\frac{8 \alpha M}{r^3}}-1} \tan ^{-1}\left(\frac{\sqrt{\sqrt{1+\frac{8 \alpha M}{r^3}}-1}}{\sqrt{2}}\right)-2 \sqrt{2} r}{3 \sqrt{\frac{r^2
   \left(\sqrt{1+\frac{8 \alpha M}{r^3}}-1\right)}{\alpha}}}\right\}\right]}\right|^2,
\end{align}
which results in a complex expression involving nested integrals with respect to $ t(r) $ and $ r_{*} $. It's important to note that the limits of integration correspond to the detector's trajectory from $ r = \infty $ to $ r = r_{+} $, the horizon of the black hole. Thus, from Eqs. (\ref{t}) and (\ref{rstar}), we derive:
  \begin{eqnarray}\label{coordtime}
      t(r)=-\int_{\infty}^{r_{+}} \frac{ \mathrm{d}r\, }{\left[1+\frac{r^2}{2\alpha}\left(1-\sqrt{1+\frac{8\alpha M}{r^3}}\right)\right]
      \sqrt{\frac{r^2}{2\alpha}\left( \sqrt{1+\frac{8\alpha M}{r^3}}-1\right)}},\ \ \ \ r_{*}&=\int_{r_{+}}^{\infty} \frac{ \mathrm{d}r\, }{1+\frac{r^2}{2\alpha}\left(1-\sqrt{1+\frac{8\alpha M}{r^3}}\right)}.
  \end{eqnarray}
  Consider now the substitution $ r = r_{+}z $, where $ \mathrm{d}r = r_{+} \mathrm{d}z $. Using this transformation of variables, we may rewrite $ t(r) $ in Eq. (\ref{coordtime}) as follows:
\begin{eqnarray}
      t(z)=-\int_{\infty}^{1}  \frac{\mathrm{d}z\, r_{+}}{\left[1+\frac{r_{+}^2z^2}{2\alpha}\left(1-\sqrt{1+\frac{8\alpha M}{r_{+}^3z^3}}\right)\right]
      \sqrt{\frac{r_{+}^2z^2}{2\alpha}\left( \sqrt{1+\frac{8\alpha M}{r_{+}^3z^3}}-1\right)}}.
  \end{eqnarray}
A further substitution of the form $x=z-1$, such that $z=x+1$, yields
\begin{eqnarray}\label{ttime}
      t(x)=\int_{0}^{\infty} \frac{ \mathrm{d}x\, r_{+}}{\left[1+\frac{r_{+}^2(x+1)^2}{2\alpha}\left(1-\sqrt{1+\frac{8\alpha M}{r_{+}^3(x+1)^3}}\right)\right]
      \sqrt{\frac{r_{+}^2(x+1)^2}{2\alpha}\left( \sqrt{1+\frac{8\alpha M}{r_{+}^3(x+1)^3}}-1\right)}}.
  \end{eqnarray}
One can follow a similar calculation for $r_{*}$, arriving at
\begin{align}
    r_{*}(x)&=\int_{0}^{\infty} \frac{ \mathrm{d}x\, r_{+}}{1+\frac{\left[r_{+}(x+1)\right]^2}{2\alpha}\left(1-\sqrt{1+\frac{8\alpha M}{\left[r_{+}(x+1)\right]^3}}\right)} .
\end{align}
After deploying all the relevant quantities in Eq. (\ref{probalpha}), we derive the following final expression for the detector excitation:
\begin{align}\nonumber
  \Gamma_{\rm exc}&= g^2 r_{+}^2\left|\int_{0}^{\infty}  \mathrm{d}x\ \ \ \exp{\left[i\nu\left\{t(x)- r_{*}(x)\right\}\right]} 
    \frac{\mathcal{G}}{\sqrt{\frac{(r_{+}[x+1])^2}{2\alpha}\left( \sqrt{1+\frac{8\alpha M}{(r_{+}\left[x+1\right])^3}}-1\right)}}\right|^2,
\end{align}
where 
\begin{eqnarray}\label{main result}
   \mathcal{G}= \exp{\left[i\omega\left\{\frac{2 r_{+}\left(x+1\right) \sqrt{\sqrt{1+\frac{8 \alpha M}{(r_{+}\left[x+1\right])^3}}-1} \tan ^{-1}\left(\frac{\sqrt{\sqrt{1+\frac{8 \alpha M}{(r_{+}\left[x+1\right])^3}}-1}}{\sqrt{2}}\right)-2 \sqrt{2} \left[r_{+}(x+1)\right]}{3 \sqrt{\frac{(r_{+}\left[x+1\right])^2
   \left(\sqrt{1+\frac{8 \alpha M}{(r_{+}\left[x+1\right])^3}}-1\right)}{\alpha}}}\right\}\right]}.
\end{eqnarray}
\end{widetext}
\begin{figure*}[t]
\centering
\includegraphics[width=\linewidth, height=13cm]{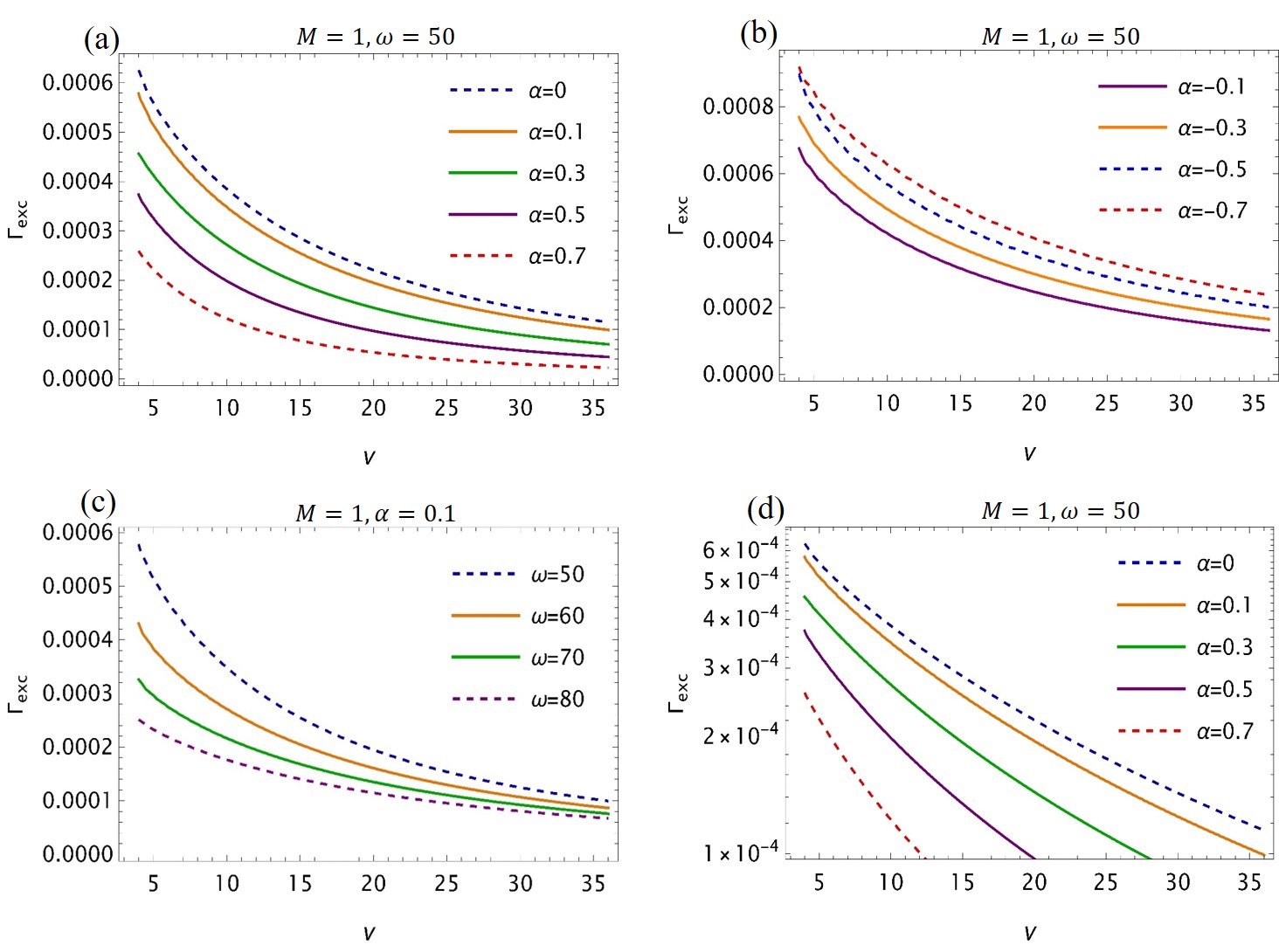}
\caption{\textcolor{black}{Impact of Gauss-Bonnet coupling $\alpha$ on  detector response $\Gamma_{\rm exc}$:(a) positive $\alpha$, and (b) negative $\alpha$, for a fixed black hole mass $M$.(c) indicates impact of detector transition frequency $\omega$ where we set $\alpha=0.1$, and (d) is the log-log plot of $\Gamma_{\rm exc}$ to  emphasize the BE-type nature of the spectrum. We chose  $g=1$ throughout.}}%
\label{PGB}%
\end{figure*}

This represents the primary outcome of our investigation. The numerical integral in \eqref{main result} is notably intricate, demanding a careful approach for its accurate computation. To achieve that, in what follows we deploy the numerical integration capabilities of the \textit{Mathematica} symbolic math package for performing all required calculations. The figures presented in Fig. \ref{PGB} were generated using optimized settings.

\section{Results and discussions}\label{discussionEGB}
 
Based on the preceding analysis, the two-level Unruh-DeWitt detector, operating in the Boulware vacuum state, registers detections while in free fall (inertial). This observation appears to challenge established field-theoretic concepts associated with the Hawking-Unruh effect. Specifically, there is no emission of Hawking radiation in the Boulware state as observed from asymptotic infinity, nor does the Unruh effect manifest for inertial detectors in the Minkowski vacuum. 
However, HBAR emission from detectors operates on different principles \cite{Scully:2017utk,Ben-Benjamin:2019opz}. While it shares similarities with Hawking radiation, such as the thermal nature of the emitted flux and the associated Bekenstein-Hawking entropy-area correspondence, there are also distinct characteristics. Notably, HBAR emission involves the evolution of field modes in pure states and includes phase correlations between them. These aspects naturally relate to the black hole information paradox \cite{Hawking:1976ra,Almheiri:2020cfm}. \\
\indent In Fig. \ref{PGB}, we present the detector excitation probability, $\Gamma_{\rm exc}$, plotted as a function of the emitted radiation frequency, $\nu$. The impact of the GB coupling parameter, $\alpha$, is depicted in Figs. \ref{PGB}(a) and \ref{PGB}(b) for positive and negative values of $\alpha$, respectively. Fig. \ref{PGB}(c) illustrates how the detector transition frequency, $\omega$, influences $\Gamma_{\rm exc}$, while Fig. \ref{PGB}(d), shown on a log-log scale, highlights the behavior of $\Gamma_{\rm exc}$ near the origin and its convergence at higher frequencies.
It is important to note that our interpretations and analyses are based on numerical estimations detailed in the preceding sections. These figures provide a comprehensive view of the radiative characteristics under consideration, elucidating the role of $\alpha$ and the detector's transition frequency in shaping $\Gamma_{\rm exc}$. \\
\indent 
From all plots, one of the prominent features observed is the thermal nature of the HBAR radiation flux, characterized by a Bose-Einstein (BE) distribution. This observation leads us to conclude that $4$D Einstein-Gauss-Bonnet (EGB) gravity does not alter the thermal nature of the flux, consistent with earlier findings \cite{Scully:2017utk,Ben-Benjamin:2019opz,Chakraborty:2019ltu,Azizi:2021qcu,Azizi:2021yto} in the context of Einstein gravity. This characteristic mirrors the thermal emission observed in Hawking radiation from pure black holes with asymptotically flat geometries.
It is noteworthy that for the so-called ``dirty'' black holes beyond the Kerr-Newman family, such as in the de Sitter case, there exists the possibility of observing a nonthermal spectrum \cite{Kastor:1993mj,Visser:2014ypa,Qiu:2019qgp,Bukhari:2022wyx}.
The detector excitation probability $\Gamma_{\rm exc}$, as observed in Fig. \ref{PGB}(a), decreases with increasing positive values of $\alpha$ and increases with negative values of $\alpha$. As previously discussed, positive $\alpha$ reduces the size of the black hole horizon [see Fig. \ref{frrgplot}(b)], leading to the conclusion that smaller black holes emit less radiation flux compared to larger ones. This reasoning can similarly be applied to negative values of $\alpha$.
It is crucial to emphasize that in the limit $\alpha \rightarrow 0$, depicted in Fig. \ref{PGB}(a), the scenario converges to that of the pure Schwarzschild black hole. \\
\indent 
The attenuation and augmentation of particle production can be conceptually grasped as follows.
Particles generated within black hole spacetimes, as in Hawking radiation, experience backreaction due to the gravitational tidal forces exerted by the black hole. This backreaction diminishes the intensity of the radiation. Tidal effects in black holes stem from their surface gravities, which are directly related to their horizon radii. Specifically, for a black hole with a horizon radius $ r_{g} $, the surface gravity varies inversely with the square of $ r_{g} $. This relationship implies that larger black holes have smaller surface gravities and correspondingly weaker tidal effects, and conversely, smaller black holes exhibit stronger tidal effects. \\
\indent 
In the context of positive $ \alpha $, the black hole horizon size decreases monotonically, leading to stronger surface gravity and tidal effects compared to a Schwarzschild black hole ($ \alpha = 0 $). Consequently, particles generated within such spacetimes experience heightened backreaction, impeding their propagation. This results in fewer particles escaping to asymptotic infinity, thereby reducing the intensity of the particle spectrum, as illustrated in Fig. \ref{PGB}(a).
Conversely, for negative values of $ \alpha $, the black hole horizon radius increases, indicating reduced backreaction and tidal forces. This condition allows more particles to escape from the black hole spacetime, leading to an enhancement in the radiation flux, as depicted in Fig. \ref{PGB}(b). This constitutes the primary finding of our study, distinguishing $ 4 $D EGB gravity from Einstein GR.\\
\indent To investigate the influence of the detector transition frequency $ \omega $ on the radiation intensity, we plot $ \Gamma_{\rm exc} $ against $ \omega $ in Fig. \ref{PGB}(c). The graphs clearly demonstrate a decrease in radiation intensity as $ \omega $ increases. This behavior aligns with the principles of the Unruh effect \cite{Unruh:1976db} and can be understood in terms of energy conservation: higher detector transition frequencies require more energy to excite the detector, resulting in a lower excitation probability, and vice versa.\\ 
\indent Furthermore, to gain insight into the spectrum's behavior at low and high frequencies ($ \nu $), we reexamined $ \Gamma_{\rm exc} $ from Fig. \ref{PGB}(a) using a log-log scale. It is evident that the spectrum exhibits a finite Bose-Einstein (BE)-type distribution near the origin where $ \nu \rightarrow 0 $. As $ \nu $ increases, the spectrum converges and exhibits a thermal tail, characteristic of a BE or Planckian distribution.
\section{possible observational imprints of HBAR?}
 Though above investigation predominantly revolves around behavior of quantum fields emanating and  propagating in curved spacetimes, it is however pertinent to consider the testable implications of this study. While saying so, we must note that there is no direct evidence  of Hawking radiation in real astrophysical settings till date. Hawking radiation  is one of the pivotal predictions of the framework of quantum field theory in curved spacetime \cite{Birrell:1982ix}. The real difficulty in making any direct observations comes from the sparsity of Hawking  flux. For example, Hawking temperature 
 \begin{eqnarray}
     T_{\rm H}=\frac{\hbar c^3}{8\pi k_{\rm B} G M}
 \end{eqnarray}
 of a solar mass ($M_\odot=10^{30}\rm kg$ ) black hole is $\sim 10^{-8}\rm K$, which means that a black hole radiating at such a temperature produces   a very tiny signal by astronomical standards when it comes to the detection.  The situation is likewise not that encouraging for Unruh effect as well. Figuratively speaking, for a body in flat spacetime to experience a temperature $1\rm K$, it needs to undergo an acceleration of about $\sim 10^{20}\rm m/s^2$ \cite{Hu:2018psq}, which obviously is beyond the threshold capacity of currently the most powerful particle accelerators on Earth. These facts are very well-known and  hence they have served as an impetus for people to look for alternative programs. In this vein, one of the notable ideas is that of  analog gravity systems \cite{Barcelo:2005fc,Braunstein:2023jpo,Jacquet:2020bar}, which have been actively utilized over the last few decades to realize quantum field-theoretic phenomena in controlled lab environments. These systems have provided valuable analogs for understanding effects such as Hawking radiation \cite{Weinfurtner:2010nu}, the Unruh effect \cite{Hu:2018psq}, and Parker particle generation in expanding spacetimes \cite{Steinhauer:2021fhb}. Most of these setups, whether based on condensed matter or quantum optical systems, are closely tied to Casimir physics involving moving boundaries. This connection is particularly relevant to our study, which is largely inspired by these concepts. Recently, condensed matter systems have been utilized to explore phenomena extending beyond particle generation in exotic backgrounds, including applications for fluid/gravity correspondence \cite{Hubeny:2011hd}. Looking forward, there is a potential for future tabletop experiments to simulate black hole horizons resembling those in $4$D EGB gravity \cite{Glavan:2019inb} or to investigate HBAR radiation scenarios \cite{Scully:2017utk}, where our findings could offer any possible insights.

\setlength{\parskip}{0cm}
    \setlength{\parindent}{0.5em}
\section{Conclusion and outlook}\label{secconclusions}
\setlength{\parskip}{0cm}
    \setlength{\parindent}{1em}
 The exploration of theories beyond Einstein gravity has evolved in parallel with general relativity (GR) itself. These modified or extended gravity theories aim to tackle fundamental cosmological issues such as cosmic acceleration, singularities, and dark matter. Among these models, Einstein-Gauss-Bonnet (EGB) gravity stands out, predicting higher-curvature corrections to the Einstein-Hilbert action. These corrections arise either in higher dimensions or through additional field couplings to the gravitational action.
 Interestingly, these contributions also emerge in the low-energy effective description of heterotic string theory. The 4D EGB theory represents a novel gravitational model that has sparked intense debate since its inception several years ago. This model predicts the presence of a Gauss-Bonnet (GB) term within 4D spacetime, which would otherwise not contribute to the latter's geometry. Its ability to provide a nontrivial contribution is achieved through a redefinition (rescalling) of the GB parameter \cite{Glavan:2019inb}. Importantly, this theory circumvents the Lovelock theorem and sidesteps Ostrogadsky instability, ensuring that the resulting gravitational dynamics remain quadratic. The theory has been scrutinized across various phenomenological fronts.\\
 \indent In this paper, we examined the quantum radiative properties of a nonrotating, uncharged black hole in $4$D Einstein-Gauss-Bonnet (EGB) gravity using a Casimir-type configuration. The black hole, surrounded by a reflecting mirror, induced accelerated field modes from the Boulware vacuum state. We analyzed the interaction of these field modes with a freely falling two-level Unruh-DeWitt detector, which exhibited characteristic clicking behavior akin to the Unruh effect.\\
 \indent The spectrum detected by the detector follows a Bose-Einstein (BE) distribution, with a notable dependence on the GB parameter $\alpha$. By examining both positive and negative values of $\alpha$, we studied their influence on the radiation intensity emitted by the detector. We observed that radiation intensity diminishes when $\alpha$ is positive. This reduction is attributed to the shrinking of the black hole size caused by positive $\alpha$. Conversely, for negative $\alpha$, we observed an increase in radiation intensity. The reduction or augmentation of the radiation flux is examined in relation to a pure Schwarzschild black hole, where the limit $\alpha \rightarrow 0$ is considered. Additionally, we observed that the transition frequency of the detector reduces the profile of particle creation due to the high energy needed for its excitation, consistent with the standard predictions of the Unruh effect. Finally, the spectrum is finite near the origin and monotonically converges at the high end of the frequency ranges, yielding the distinctive thermal tail characteristic of a Bose-Einstein or Planckian distribution.\\
\indent Our work provides an opportunity to explore various aspects of 4D EGB gravity by incorporating different energy-matter distributions around the simplest black hole model possible. Moreover, exploring other types of detector-field couplings could yield valuable insights into the nature of field configurations within the context of 4D EGB gravity. These and similar questions constitute promising extensions of this work, which we plan to pursue in the future.

\bibliographystyle{apsrev4-1}
\bibliography{masood.bib}
\end{document}